\documentclass[letterpaper, 10 pt, conference]{ieeeconf}  

\usepackage{amsmath,amsfonts,bm}

\usepackage{algorithmic}
\usepackage{array}
\usepackage[caption=false,font=normalsize,labelfont=sf,textfont=sf]{}
\usepackage{textcomp}
\usepackage{stfloats}
\usepackage{amsthm} 
\usepackage{url}
\usepackage{verbatim}
\usepackage{graphicx}
\usepackage{hyperref}
\usepackage{graphics}
\usepackage{siunitx}
\usepackage{colortbl, soul}
\definecolor{lightgray}{rgb}{0.753, 0.753, 0.753}
\usepackage{amsmath}               
  {
      \theoremstyle{plain}
      
  }

\hypersetup{
	colorlinks=true,       		
	linkcolor=black,          	
	citecolor=black,        		
	filecolor=black,      		
	urlcolor=black	
}

\usepackage{hyperref}
\hypersetup{
    pdfauthor={},
    pdftitle={},
    pdfsubject={},
    pdfkeywords={}
}

\usepackage[ruled]{algorithm2e}
\usepackage{subfiles}
\usepackage{graphicx}
\usepackage{subfigure}
\usepackage{xstring}
\usepackage{tabularx}
\usepackage{booktabs} 
\usepackage{float} 

\usepackage{multirow}
\usepackage{tikz}
\usetikzlibrary{fadings}
\usetikzlibrary{patterns}
\usetikzlibrary{shadows.blur}
\usetikzlibrary{shapes}

\IEEEoverridecommandlockouts                              

\title{\LARGE \bf Contraction Metric Based Safe Reinforcement Learning Force Control for a Hydraulic Actuator with Real-World Training}

\author{Lucca~Maitan, Lucas Toschi, Cícero~Zanette, Elisa~G.~Vergamini, Leonardo~F.~Santos, \\   and Thiago~Boaventura}

\begin{document}
\maketitle

\thispagestyle{empty}
\pagestyle{empty}


\begin{abstract}
Force control in hydraulic actuators is notoriously difficult due to strong nonlinearities, uncertainties, and the high risks associated with unsafe exploration during learning. This paper investigates safe reinforcement learning (RL) for hydraulic force control with real-world training using contraction-metric certificates. A data-driven model of a hydraulic actuator, identified from experimental data, is employed for simulation-based pretraining of a Soft Actor-Critic (SAC) policy that adapts the PI gains of a feedback-linearization (FL) controller. To reduce instability during online training, we propose a quadratic-programming (QP) contraction filter that leverages a learned contraction metric to enforce approximate exponential convergence of trajectories, applying minimal corrections to the policy output. The approach is validated on a hydraulic test bench, where the RL controller is trained directly on hardware and benchmarked against a simulation-trained agent and a fixed-gain baseline. Experimental results show that real-hardware training improves force-tracking performance compared to both alternatives, while the contraction filter mitigates chattering and instabilities. These findings suggest that contraction-based certificates can enable safe RL in high-force hydraulic systems, though robustness at extreme operating conditions remains a challenge.
\end{abstract}


\section{Introduction}

Force control is a fundamental capability in robotics, as it enables safe and precise interaction with the environment by regulating the exchange of forces during contact. In contrast to position control, force control allows robots to manipulate delicate objects, adapt to changing interaction conditions, and safely collaborate with humans \cite{siciliano2008}. Effective physical interaction typically requires the coordinated regulation of both force and motion. Rather than treating these variables independently, impedance control shapes the dynamic behavior of the system by prescribing desired stiffness, damping, and inertia through an internal force control loop \cite{hogan1984impedance}. While this strategy promotes compliant and adaptive behavior, particularly in uncertain or dynamic environments, its overall performance is strongly influenced by the quality of the internal force control loop. As a result, achieving accurate and robust force control remains especially challenging under such conditions \cite{boaventura2012role}.

Consequently, the development of high-performance force controllers is of central importance. Among electric, pneumatic, and hydraulic actuation technologies, hydraulic actuators are particularly attractive due to their high power-to-weight ratio, which enables compact designs capable of delivering large forces. Despite the inherent nonlinearities and control challenges associated with hydraulic systems, they have been widely adopted in high-performance robotic platforms \cite{Boaventura22}. The \emph{HyQ} quadruped robot exemplifies the advantages of hydraulic actuation in legged locomotion \cite{semini2010hyq}, as do bipedal systems such as ATLAS \cite{nelson2012petman}, and SARCOS \cite{atkeson00sarcos}.

A number of recent works have proposed advanced force control strategies for hydraulic actuators. For instance, \cite{sheng2016hybrid,zhao2019robust} employed linearized hydraulic models combined with disturbance observers. In \cite{nakkarat2009observer}, a disturbance observer was integrated with a backstepping-based force controller for load simulation applications. More recently, \cite{fang2025adaptive} introduced a robust adaptive control approach capable of estimating uncertain nonlinear nominal parameters for force control in a hydraulic robot leg.

In parallel, reinforcement learning (RL) has emerged as a promising alternative to classical control techniques. By learning control policies directly through interaction, RL reduces reliance on explicit system modeling and can offer improved adaptability in complex and uncertain environments \cite{perrusquia2019position}. Several studies have combined RL with conventional control structures in hydraulic systems, including RL-based gain adaptation of PID controllers \cite{yao2023data} and disturbance estimation using actor–critic architectures \cite{yao2022model}. In \cite{gao2024reinforcement}, multiple RL strategies were implemented and compared for position control of a hydraulic inverted pendulum. Additionally, RL-based controllers have been successfully applied to hydraulic machinery handling heavy materials and excavation tasks \cite{spinelli2024reinforcement, gruetter2025towards}. To address the challenges of deploying learned policies on real hardware, techniques such as dynamics and domain randomization have been widely adopted to improve sim-to-real transfer \cite{peng2018sim,tobin2017domain}.

One area of research currently being studied is the utilization of real-life hardware training to bypass sim-to-real transfer issues of policies. However, safety remains a central obstacle for real-robot training, as naive exploration can violate physical limits, damage hardware, or endanger people. As a result, most prior reinforcement learning studies in robotics restrict training to simulation and only perform deployment on the real platform. Safe RL frameworks have been proposed to mitigate risks during learning via constrained optimization, formal-methods shielding, and control-theoretic safety layers such as control barrier functions \cite{achiam2017constrained,alshiekh2018safe,cheng2019end}. Beyond these, contraction theory provides trajectory-level robustness certificates by guaranteeing exponential convergence between trajectories. Contraction metrics and related control designs yield intrinsic and convex conditions that can be integrated with learning. Recent research has also focused on learning contraction metrics with neural networks, enabling stability certificates even in high-dimensional nonlinear systems \cite{sun2021learning}. Surveys further highlight the role of contraction-based and other learned certificates in safe control for robotics \cite{dawson2023safe}. 

Despite this progress, implementations of safe RL that train policies directly on real robotic hardware remain scarce, particularly in systems with highly nonlinear dynamics such as hydraulics. Hydraulic actuators can generate extremely large forces and power densities, which amplifies the risk of unsafe exploration and makes the development of online learning strategies far more critical. Demonstrating that safe online training is feasible on such platforms therefore represents a significant step beyond existing work, which has largely remained in simulation or relied on conservative safety margins during deployment. This motivates the present study, which combines learning-based force control, reinforcement learning, and contraction-metric safety certificates in a hydraulic actuation setting.

Our approach is related to Lyapunov and control barrier function (CBF) methods. The main difference is the focus of the guarantee that we wish to enforce. Lyapunov and CBF methods are typically designed to keep the system near a fixed operating point or within a predefined safe region \cite{dawson2023safe}. In contrast, contraction-based analysis is naturally suited to reference-tracking problems, as it certifies convergence between trajectories rather than stability with respect to a single equilibrium or constraint set. This is particularly relevant in our setting, where time-varying force references and online policy updates cause the desired behavior to evolve during training. Enforcing contraction, therefore, directly targets tracking-related instabilities that arise as the reference and controller gains change over time. 

Therefore, to the best of our knowledge, this paper makes three contributions:
(i) Safe real-world training of an RL policy on a hydraulic system with contraction-metric safety certificates, demonstrating the feasibility of online learning under strong nonlinearities and uncertainties;
(ii) Application of contraction theory to hydraulic force control, where we project the contraction condition to the force-tracking dimension and embed it into a lightweight QP filter for real-world use; and
(iii) Development and hardware validation of a contraction-based safety filter that enforces learned certificates online while respecting actuator limits, reducing instabilities during RL training, and complementing simulation-to-real pretraining.

\section{Hydraulic Force Dynamics}
\label{sec:hyd}

We follow a hydraulic force model previously introduced for quadruped robots \cite{BoaventuraThesis}, which shares a similar hydraulic platform to the experimental test bench used in this work (see Section~VI). For brevity, we present only the final form of the model in~\eqref{eq:q13}, which serves as the foundation for the model-based controller integrated into the reinforcement learning framework. The corresponding parameters are listed in Table~\ref{tab:param}.

\begin{table}[htb]
\centering
\caption{Hydraulic model parameters.}
\resizebox{0.8\linewidth}{!}{%
\begin{tabular}{@{}ccc@{}}
\toprule
\textbf{Symbol} & \textbf{Parameter} & \textbf{Value} \\ \midrule
$\beta_{e}$  & Fluid bulk modulus [\unit{\pascal}] & \num{1.34e9} \\
$A_{p}$ & Chamber A area [\unit{\square\meter}] & \num{2e-4}\\
$p_{a}$ & Pressure in chamber A [\unit{\mega\pascal}] & Measured \\
$p_{b}$ & Pressure in chamber B [\unit{\mega\pascal}] & Measured \\
$p_{s}$ & Source pressure [\unit{\mega\pascal}] & 16 \\ 
$p_{t}$ & Tank pressure [\unit{\mega\pascal}] & 0 \\
$L_{c}$ & Max piston displacement [\unit{\meter}] & 0.08 \\
$q_{a}, q_{b}$ & Chamber flows [\unit{\cubic\meter\per\second}] & Measured \\
$f_{h}$ & Hydraulic force [\unit{\newton}] & Measured \\
$\alpha$ & Chamber area ratio & 0.609 \\
$x_{p}$ & Piston displacement [\unit{\meter}] & Measured \\
$v_{a}, v_{b}$ & Chamber volumes [\unit{\cubic\meter}] & Measured \\
$u$ & Input valve current [\unit{\ampere}] & Measured \\
$q_{n}$ & Valve nominal flow  [\unit{\cubic\meter\per\second}] & \num{1.67e-4}\\
$u_{n}$ & Valve nominal current  [\unit{\milli\ampere}] & \num{50}\\
$\Delta p_{n}$ & Valve nominal pressure drop [\unit{\mega\pascal}]& \num{7} \\
 \bottomrule
\end{tabular}}
\label{tab:param}
\end{table}

\begin{equation}
\dot f_{h} = h(x_{p},\dot x_{p}) + g(x_{p},P)u,
\label{eq:q13}
\end{equation}

\noindent with
\begin{equation}
h(x_{p},\dot x_{p}) = -\beta_e A_{p}^{2}\left (\frac{1}{v_{a}} + \frac{\alpha^{2}}{v_{b}} \right)\dot x_{p},
\label{eq:q14}
\end{equation}

\begin{equation}
    g(x_{p},P) = 
    \begin{cases}
        \beta_{e}K_{v}A_{p} \left(\dfrac{\sqrt{p_{s} - p_{a}}}{v_{a}} - \dfrac{\alpha\sqrt{p_{b} - p_{t}}}{v_{b}} \right), \ u \ge 0 \\
        \beta_{e}K_{v}A_{p} \left(\dfrac{\sqrt{p_{a} - p_{t}}}{v_{a}} - \dfrac{\alpha\sqrt{p_{s} - p_{b}}}{v_{b}} \right), \ u \le 0,
    \end{cases}
    \label{eq:q15}
\end{equation}

\noindent where $P = (p_{a},p_{b},p_{s},p_{t})$ collects the chamber and pump pressures, and $K_{v}$ is the valve gain, obtained by linearizing the valve flow around its nominal operating point:

\begin{equation}
K_{v} = \frac{q_{n}}{u_{n}\sqrt{\tfrac{\Delta p_{n}}{2}}}.
\label{eq:q4}
\end{equation}

Although~\eqref{eq:q13} captures the essential force dynamics, practical actuators are also affected by viscous and Coulomb friction, as well as internal and valve leakage. These phenomena slow down the response and reduce steady-state accuracy~\cite{singh2022effect}. For simplicity, their combined influence is modeled as a lumped disturbance term \( d \):

\begin{equation}
\dot f_{h} = h + g u + g d.
\label{eq:model_cert}
\end{equation}

To further account for modeling errors, we introduce uncertainty scalars $C_1$ and $C_2$ applied to~\eqref{eq:q14} and~\eqref{eq:q15}, respectively:

\begin{equation}
\dot f_{h} = C_{1}h + C_{2}g u + g d.
\label{eq:model_uncert}
\end{equation}

This formulation highlights the main sources of uncertainty: the valve gain $K_{v}$, derived from linearization, and the fluid bulk modulus $\beta_e$, which is inherently variable. Since both appear linearly in the dynamics, they can be represented by scalar multipliers. Nevertheless, because $d$, $C_1$, and $C_2$ are unknown, model~\eqref{eq:model_uncert} cannot by itself guarantee precise prediction of the system’s response. Relying exclusively on this model can therefore hinder controller performance, motivating the integration of reinforcement learning to improve tracking and robustness.

\section{Hydraulic Model Learning}
As stated in Section \ref{sec:hyd}, \eqref{eq:model_uncert} does not represent the system with enough accuracy, however, a good approximation of the hydraulic model is necessary for the implementation of the developments of this paper as detailed in Section \ref{sec:ccm}. This following section describes the application of supervised learning strategies to learn a more accurate model for the system.   
\label{sec:model_learn}

\subsection{Model Learning}

\begin{figure}[b]
  \centering
  \includegraphics[width=0.8\linewidth]{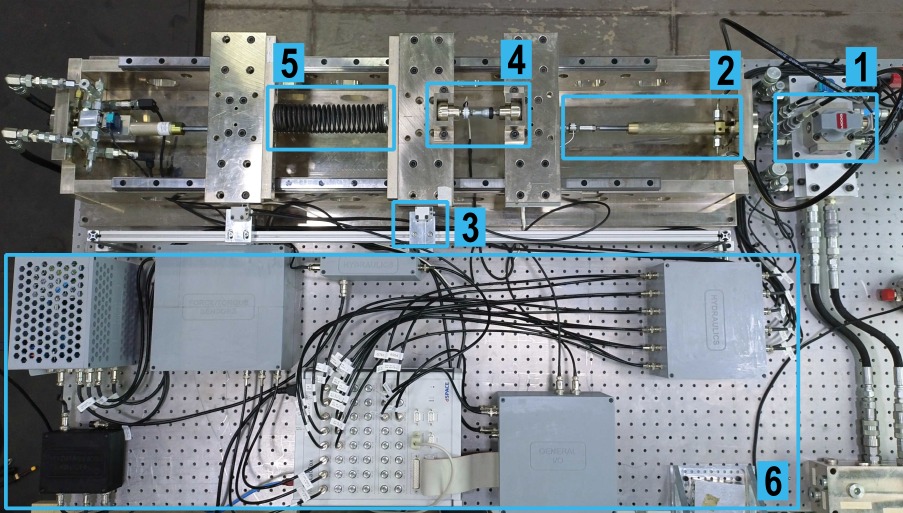}
  \caption{Experimental setup used for controller testing and environment model learning. 1 - Servo valve. 2 - Cylinder. 3 - Incremental encoder. 4 - Load cell. 5 - Spring with stiffness \SI{20000}{\newton\per\meter}. 6 - Electronics.} 
  \label{Graph6}
\end{figure}

The system under study is a hydraulic test bench experimental hardware, illustrated in \autoref{Graph6}. The hydraulic actuation is realized through a Moog G761/-761-V10JOXM4VPM servo valve coupled with a Hoerbiger LB6 1610 0080 double-action cylinder, the same configuration used in the \emph{HyQ} robot. The model parameters are summarized in Table~\ref{tab:param}. 

Our control objective is to regulate the interaction force $f_l$, measured by a load cell at the end-effector. This interaction occurs through contact with an external spring environment. The actuator force $f_h$, applied prior to $f_l$, is also directly measured by a dedicated load cell.

As discussed in Section~\ref{sec:hyd}, a purely analytic model is insufficient for high-fidelity simulation. To enable reinforcement learning (RL) policy training in simulation, we therefore learned a more accurate dynamics model using a Multilayer Perceptron (MLP). The network predicts the system evolution at each time step:

\begin{equation}\label{eq:state}
\dot x_{t} = f(x_t,u_t),
\end{equation}

\noindent where $u_t$ is the applied valve current and $x_t$ is the state vector:

\begin{equation}\label{eq:state_def}
x_t = [f_{h}, \dot f_{h}, f_{l}, \dot f_{l}, p_{a}, p_{b}, x_p, \dot x_p].
\end{equation}

To train this model, we collected a dataset from the real system by executing randomly generated reference trajectories with a stable force controller. The MLP was then optimized to minimize the following loss:

\begin{equation}\label{eq:loss}
\mathcal{L}(x_t) = \frac{1}{N}\sum_{i=1}^{N}\Bigg(\sum_{h=0}^{H} \delta_{t+h+1}\Bigg)^{2},
\end{equation}

\noindent where $\delta$ denotes the error between predicted and measured states. Early experiments with single-step prediction ($H=1$) resulted in poor accuracy, as the model did not capture longer-term dynamics. To overcome this, we used a multi-step prediction horizon $H$, propagating model outputs recursively and computing the cumulative error across predicted states. This strategy significantly improved performance. Training was performed on normalized data using an NVIDIA RTX A2000 GPU (12 GB). The main training hyperparameters are summarized in Table~\ref{tb:training}.

\begin{table}[b]
\centering
\caption{Hydraulic model learning parameters.}
\label{tb:training}
\resizebox{0.67\linewidth}{!}{%
\begin{tabular}{rp{0.26\columnwidth}}
\toprule
\textbf{Parameter} & \textbf{Value} \\ \midrule\midrule
\rowcolor{lightgray}
Epochs & $1000$\\
Batch Size & $8192$\\
\rowcolor{lightgray}
Hidden layers & $2$\\
Hidden size & $32$\\ 
\rowcolor{lightgray}
Learning rate & $0.001$\\
Prediction horizon $H$ & $70$\\
\rowcolor{lightgray}
Activation & ReLU\\
\bottomrule
\end{tabular}}
\end{table}

\subsection{Model Validation}

Table~\ref{tb:rmse_model} reports the Root Mean Squared Error (RMSE) of normalized predicted states from \eqref{eq:state_def}, comparing our trained model with the analytic formulation of~\eqref{eq:q13}. The learned model achieves errors more than two orders of magnitude smaller, demonstrating its ability to accurately reproduce the system dynamics and significantly outperform the analytic model.

\begin{table}[H]
\centering
\caption{Normalized model prediction RMSE.}
\label{tb:rmse_model}
\resizebox{\linewidth}{!}{%
\begin{tabular}{|c|l|l|l|l|l|l|l|l|}
\hline
\multicolumn{1}{|l|}{} & \multicolumn{1}{c|}{$f_h$}  & \multicolumn{1}{c|}{$\dot{f_h}$}  & \multicolumn{1}{c|}{$f_l$}  & \multicolumn{1}{c|}{$\dot{f_l}$}  & \multicolumn{1}{c|}{$p_a$}  & \multicolumn{1}{c|}{$p_b$}  & \multicolumn{1}{c|}{$x_p$}  & \multicolumn{1}{c|}{$\dot{x_p}$}     \\ \hline
Trained model   & 0.0006 & 0.0000 & 0.0009 & 0.0000 & 0.0004 & 0.0001 & 0.0031 & 0.0000 \\ \hline
Analytic model  & 0.5658 & 0.0362 & 0.9094 & 0.1238 & 0.0958 & 0.1535 & 0.1223 & 0.1375 \\ \hline
\end{tabular}}
\end{table}

\section{Baseline Feedback Linearization Controller}
\label{sec:fl}

As a baseline for reinforcement learning (RL) policy training, we employ a feedback linearization (FL) controller designed for the nonlinear force dynamics, following the approach used for hydraulic quadruped robots in~\cite{BoaventuraThesis}. Assuming the nominal force dynamics of~\eqref{eq:q13}, the control law is:

\begin{equation}
u = \frac{\dot{f_{r}} + K_{p}e + K_{i}\int e \, dt}{g} - \frac{h}{g},
\label{eq:q24}
\end{equation}

\noindent where $f_{r}$ is the desired reference force, $e = f_{r} - f_{h}$ is the tracking error, and $K_{p}$ and $K_{i}$ are the proportional and integral gains of the inner PI controller and the unknown parameters are considered as $C_1 = 1$, $C_2 = 1$, and $d = 0$. Under ideal conditions (perfect model knowledge and no disturbances), substitution of~\eqref{eq:q24} into~\eqref{eq:q13} yields the closed-loop error dynamics:

\begin{equation}
\dot e = -K_{p} e - K_{i}\int e \, dt,
\label{eq:q25}
\end{equation}

\noindent which is linear and asymptotically stable.

In practice, however, the force dynamics of~\eqref{eq:model_uncert} are imprecise due to uncertainties such as valve gain variability, bulk modulus fluctuations, and unmodeled disturbances. Consequently, exact cancellation of the nonlinearities cannot be guaranteed, and controller performance depends heavily on the chosen $K_{p}$ and $K_{i}$ values. This motivates the use of RL to adapt and optimize the PI gains online, as described in Section~\ref{sec:policy_learn}.

We acknowledge that this is a simple nonlinear control structure and that comparisons with more robust controllers would provide additional perspective, and leave this as a direction for future work. Our objective in this work is to highlight the potential of RL to enhance controller performance, especially RL trained in real-world hardware.


\section{Contraction Metric Certificate}
\label{sec:ccm}
When training RL policies directly on the real system, exploratory actions can transiently destabilize the closed loop, risking hardware and data quality. Our goal is to endow the baseline controller + RL policy with a safety certificate that (i) detects impending loss of stability and (ii) minimally corrects the input to preserve stability without overwriting nominal performance. Contraction theory is well suited for this: it certifies exponential convergence of neighboring trajectories independent of a specific reference, enabling a lightweight quadratic-program (QP) filter that guards the system while RL adapts the controller.

\subsection{Contraction Metric Overview}
\label{sec:ccm_1}

Contraction theory provides a framework to verify closed-loop stability by analyzing the rate of convergence between neighboring trajectories of a system. If the distance between any two trajectories decreases exponentially, the system is said to be contracting and thus stable for arbitrary reference trajectories~\cite{lohmiller1998contraction}. Formally, a closed-loop controller ensures:

\begin{equation}\label{eq:cont_require}
\|x(t) - x_r(t)\| \le K e^{-\lambda t} \|x(0) - x_r(0)\|, \quad \forall t \ge 0,
\end{equation}

\noindent where $x$ is the system state, $x_r$ is the reference trajectory, $\lambda > 0$ is the contraction rate, and $K \geq 1$ is a bounding constant. 

This condition holds if there exists a state-dependent, symmetric, positive-definite matrix $M(x)$ (a contraction metric) such that for the differential displacement $\delta x$ between neightboring trajectories, with candidate Lyapunov function $V = \delta x^\top M(x) \delta x$, the following inequality is satisfied:

\begin{equation}\label{eq:cont_require_2}
\dot V \le -2\lambda \, \delta x^\top M(x) \delta x.
\end{equation}

For a nonlinear control-affine system
\begin{equation}\label{eq:cont_require_3}
\dot x = f(x,u),
\end{equation}

\noindent with input $u$, the differential dynamics can be written as:

\begin{equation}\label{eq:cont_require_41}
\dot{\delta x} = A \delta x + B \delta u,
\end{equation}

\noindent where $A = \tfrac{\partial f}{\partial x}$ and $B = \tfrac{\partial f}{\partial u}$. If the input variation is given by $\delta u = K \delta x$, then:

\begin{equation}\label{eq:cont_require_4}
\dot{\delta x} = (A + BK)\delta x.
\end{equation}

The derivative of $V$ is:

\begin{equation}\label{eq:cont_require_5}
\dot V = \dot{\delta x}^\top M \delta x + \delta x^\top \dot M \delta x + \delta x^\top M \dot{\delta x}.
\end{equation}

Substituting~\eqref{eq:cont_require_4} into~\eqref{eq:cont_require_5} and enforcing~\eqref{eq:cont_require_2}, the contraction condition can be expressed as:

\begin{equation}\label{eq:cont_require_6}
\delta x^\top \big(\dot M + \mathrm{sym}(M(A + BK)) + 2\lambda M\big)\delta x \le 0,
\end{equation}

\noindent where $\mathrm{sym}(X) = X + X^\top$. A more detailed derivation of the contraction metric certificate from \eqref{eq:cont_require} to \eqref{eq:cont_require_6} is provided in~\cite{manchester2017control}.

\subsection{Contraction Metric Filtering}
\label{sec:ccm_2}

Condition~\eqref{eq:cont_require_6} allows analysis of stability. Similar to how Lyapunov functions can be used with quadratic programming (QP) filters to enforce stability~\cite{dawson2022safe}, we adopt a contraction-based filtering strategy. The control input is augmented as:

\begin{equation}\label{eq:cont_require_7}
u = u_{nom} + \Delta u,
\end{equation}

\noindent where $u_{nom}$ is the nominal controller or policy output, and $\Delta u$ is a corrective term computed to enforce contraction. Substituting into~\eqref{eq:cont_require_41}:

\begin{equation}\label{eq:cont_require_8}
\dot{\delta x} = (A + BK)\delta x + B \Delta u.
\end{equation}

This yields:
\begin{multline}\label{eq:cont_require_9}
\dot V = v^\top \dot M v + v^\top  \big((A + BK)^\top M + M(A + BK)\big)v \\ + 2 v^\top M B \Delta u,
\end{multline}

\noindent where we denote $\delta x$ as $v$ for compactness. Enforcing the contraction condition gives:

\begin{equation}\label{eq:cont_require_10}
2 v^\top M B \Delta u \le -v^\top \big(\dot M + \mathrm{sym}(M(A + BK)) + 2\lambda M\big)v.
\end{equation}

In the system, $x$ includes several states beyond the controlled force $f_l$ (see~\eqref{eq:state}). Since the objective is force tracking, we project $\delta x$ onto the force error dimension only:

\begin{equation}\label{eq:v}
v = [0, 0, e, 0, 0, 0, 0, 0], \quad e = f_r - f_l.
\end{equation}

Thus, stability enforcement reduces to solving the following QP:

\begin{align}
\min_{\Delta u \in \mathbb{R}} \quad & \|\Delta u\| \\
\text{s.t.} \quad & a \Delta u \le b,
\end{align}

\noindent with
\begin{align}
a &= 2 v^\top M B, \label{eq:cont_require_12} \\
b &= -v^\top \big(\dot M + \mathrm{sym}(M(A + BK)) + 2\lambda M\big)v. \label{eq:cont_require_13}
\end{align}

Choosing a small $\lambda$ ensures corrections are applied only when $u_{nom}$ risks instability, avoiding unnecessary overrides of otherwise stable inputs.

\subsection{Contraction Metric Learning}
\label{sec:ccm_3}

Applying the filter requires knowledge of $M(x)$, which is generally nontrivial. Recent works~\cite{sun2021learning,dawson2023safe} demonstrate that $M(x)$ can be approximated via supervised learning. We adopt a similar strategy, training a neural network to output positive definite matrices that approximate the contraction metric.

Training data were generated using the learned hydraulic model described in \autoref{sec:model_learn}, simulating trajectories under both stable and unstable controllers. An MLP was trained to minimize:

\begin{equation}\label{eq:loss_metric}
\mathcal{L}_{M}(x_t) = \sum_{h=0}^{H_m} \big(W_l C_l + R_l\big),
\end{equation}

\noindent where $C_l$ is the contraction loss:

\begin{equation}
C_l =
\begin{cases}
\mathrm{ReLU}(M_c), & \text{stable trajectory}, \\
\mathrm{ReLU}(-M_c), & \text{unstable trajectory},
\end{cases}
\label{eq:q152}
\end{equation}

\noindent with $M_c$ defined from \autoref{eq:cont_require_6}. The intuition is that the expression in \autoref{eq:cont_require_6} tends to take negative values for trajectories labeled as stable and positive values for trajectories labeled as unstable, and the loss encourages this separation. The weighting factor $W_l$ was set to $100$. Regularization $R_l$ enforces well-conditioned, positive-definite metrics:

\begin{equation}
\begin{split}
R_l &= \lambda_F \, \lVert M(x_t) \rVert_F
+ \lambda_{\mathrm{tr}} \, \operatorname{tr}(M(x_t)) \\
&\quad - \lambda_{\log\det} \, \log \det\!\big(M(x_t) + \varepsilon I_n\big),
\end{split}
\label{eq:q153}
\end{equation}

\noindent with $\varepsilon = 10^{-6}$ and $\lambda_F = \lambda_{\mathrm{tr}} = \lambda_{\log\det} = 10^{-3}$.
The Frobenius-norm term limits the overall magnitude of the contraction metric and prevents the optimization from satisfying the contraction inequality through simple scaling. The trace regularizer further restricts the growth of the metric by penalizing the sum of its eigenvalues, discouraging excessively stiff or overly conservative solutions. The log-determinant term enforces positive definiteness of the learned metric by penalizing eigenvalues approaching zero, with the offset $\varepsilon$ included to avoid numerical issues during training.

The weighting factor $W_l$ was chosen to be significantly larger than the regularization coefficients so that the contraction loss dominated the optimization, while the regularizers acted only to maintain numerical conditioning. In practice, this choice led to stable training.

\begin{figure}[htb]
  \centering
  \includegraphics[width=1\linewidth]{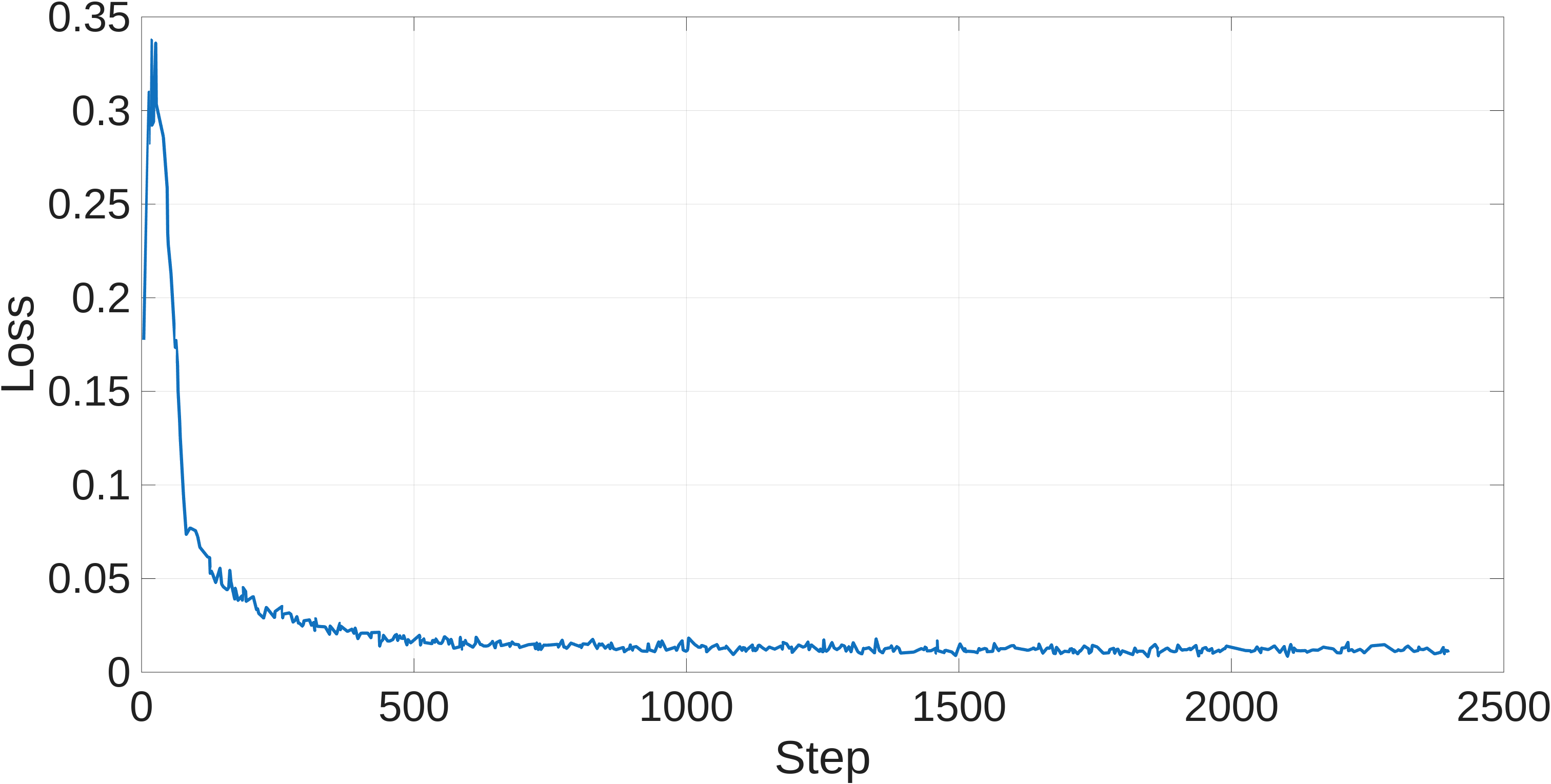}
  \caption{Contraction metric loss evolution during training} 
  \label{metric_loss}
\end{figure}

As with model learning, the loss is evaluated over a horizon $H_m$ to capture trajectory behavior. The Jacobians $A$, $B$, and $K$ are computed via finite differences at each step, which leads to very slow training, seeing as the trained model network had to be called several times for each calculation of the certificate. Due to this slow training, a low number of epochs was utilized during training as shown in Table \ref{tb:training_metric}. Training was accelerated on an NVIDIA RTX A2000 GPU (12~GB). Table~\ref{tb:training_metric} lists the hyperparameters, and Figure \ref{metric_loss} shows the evolution of the loss function \eqref{eq:loss_metric} until convergence.

\begin{table}[hb]
\centering
\caption{Contraction Metric learning parameters.}
\label{tb:training_metric}
\resizebox{0.67\linewidth}{!}{%
\begin{tabular}{rp{0.26\columnwidth}}
\toprule
\textbf{Parameter} & \textbf{Value} \\ \midrule\midrule
\rowcolor{lightgray}
Epochs & $3$\\
Batch Size & $1024$\\
\rowcolor{lightgray}
Hidden layers & $2$\\
Hidden size & $64$\\ 
\rowcolor{lightgray}
Learning rate & $0.001$\\
Simulation horizon $H_m$ & $10$\\
\rowcolor{lightgray}
Activation & ReLU\\
Optimizer & Adam\\
\bottomrule
\end{tabular}}
\end{table}

\subsection{Learned Contraction Metric Validation}
\label{sec:ccm_4}

\begin{figure}[htb]
  \centering
  \includegraphics[width=1\linewidth]{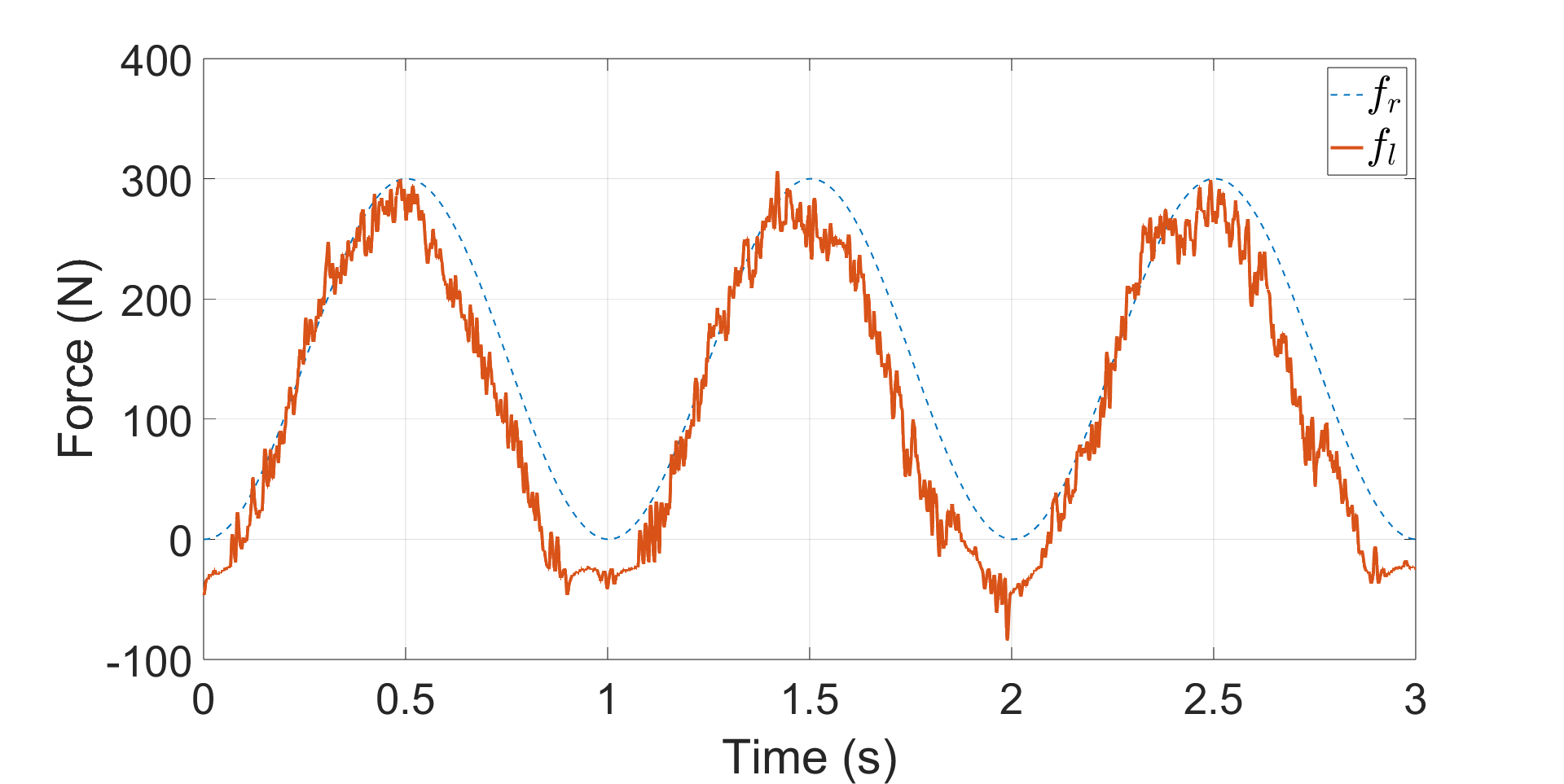}
  \caption{Force tracking with randomized PI gains using QP contraction filtering.} 
  \label{unstable_track}
\end{figure}

We validated the learned contraction metric by applying the QP filter (Section~\ref{sec:ccm_2}) on the hardware test bench. The filter was combined with the FL controller~\eqref{eq:q24}, where $K_p$ and $K_i$ were randomized with values between -40 and 40 for $K_p$ and -5 and 5 for $K_i$. This controller is described as 'Validate Unstable' in Table \ref{tb:val_cont}. Without filtering, the system became unstable; with the contraction filter ($\lambda = 0.1$), stability was maintained (see Figure \ref{unstable_track}). We also evaluated the contraction filter when utilizing a known stable controller with $K_p = 90$, and $K_i = 15$ to verify it would not lead to instability or alteration in the control action when using the same low $\lambda = 0.1$ value ('Validate Stable' in Table \ref{tb:val_cont}). The data describing how the filter affected each situation can be observed in Table \ref{tb:val_cont}, in which the metrics are presented as follows:

\begin{itemize}
  \item $\gamma_1$: Percentage mean of the value of $\Delta u$ with regard to the maximum output accepted by our hardware.
  
  \item $\gamma_2$: Percentage of time steps the value of the certificate \eqref{eq:cont_require_6} was verified before filtering

  \item $\gamma_3$: Mean value of \eqref{eq:cont_require_6}  before filtering
  
  \item $\gamma_4$: Percentage of time steps the value of the certificate \eqref{eq:cont_require_6} was verified after filtering
  
  \item $\gamma_5$: Mean value of \eqref{eq:cont_require_6}  after filtering
\end{itemize}

\begin{table}[H]
\centering
\caption{Contraction Filter Validation Results}
\label{tb:val_cont}
\resizebox{0.8\linewidth}{!}{%
\begin{tabular}{|c|lllll|}
\hline
\multicolumn{1}{|l|}{}          & \multicolumn{5}   {c|}{Metrics}
\\ \hline
\multicolumn{1}{|l|}{} & \multicolumn{1}{c|}{$\gamma_1$}  & \multicolumn{1}{c|}{$\gamma_2$}  & \multicolumn{1}{c|}{$\gamma_3$} & \multicolumn{1}{c|}{$\gamma_4$} & \multicolumn{1}{c|}{$\gamma_5$}     \\ \hline
Validate Stable                              & \multicolumn{1}{l|}{0} & \multicolumn{1}{l|}{0.14 \%} & \multicolumn{1}{l|}{-0.0528} & 
\multicolumn{1}{l|}{0.07 \%} & 
\multicolumn{1}{l|}{-0.0528} \\ \hline
Validate Unstable                              & \multicolumn{1}{l|}{0.3 \%} & \multicolumn{1}{l|}{74.75 \%} & \multicolumn{1}{l|}{0.2419} & 
\multicolumn{1}{l|}{37.50 \%} & 
\multicolumn{1}{l|}{-0.0546} \\ \hline
\end{tabular}}
\end{table}

The results shown in Table \ref{tb:val_cont} demonstrate how the QP filter did not cause any changes in the system performance when using a stable controller, as the value of $\Delta u$ did not change during the experiment as indicated by the metric $\gamma_1$ and the value of our contraction certificate \eqref{eq:cont_require_6} was already verified before and after filtering, as shown by metrics $\gamma_2$, $\gamma_3$, $\gamma_4$, and $\gamma_5$. When it comes to the unstable controller, we can see that our contraction certificate \eqref{eq:cont_require_6} detected instability before filtering, as shown by metrics $\gamma_2$, and $\gamma_3$, and that after filtering, contraction was ensured when observing $\gamma_4$, and $\gamma_5$. Although the value of $\gamma_1$ of $0.3 \%$ may seem low for the unstable controller validation, the high power of the hydraulic system makes it so that even low inputs have large power capabilities; therefore, the value $\Delta u$ was significant for the performance of the system. Contraction filter tests were also performed using strictly negative and positive random gains $K_p$ and $K_i$ for unstable controllers, and stability remained in these cases as well.

However, when controller gains exceeded large values ($K_p > 100$), the filter was insufficient to prevent instability. We attribute this to noise sensitivity at high control bandwidths: the learned metric was trained under limited noise conditions and thus generalized poorly to high-gain, noisy regimes. Another possible explanation for the difficulties of handling higher unstable gains is that the system became too fast for the filter to correct with the given implementation frequency of 100 Hz. Due to this, it was necessary to reduce the maximum values that $K_p$ and $K_i$ could achieve during training to the values shown in Table \ref{tb:training_agent}. Despite this limitation, the contraction filter significantly reduced oscillations and chattering during RL training compared to training without filtering. We anticipate that incorporating noise augmentation during metric training will mitigate these issues in future work.

\section{Real Hardware Policy Learning Description}
\label{sec:policy_learn}

We formulate the control problem as a Markov decision process (MDP), where the agent has direct access to the system state $x_t$ through measurements. The reinforcement learning (RL) strategy adopted is the Soft Actor-Critic (SAC) algorithm~\cite{haarnoja2018soft}, implemented using the \texttt{skrl} library~\cite{serrano2023skrl} and the Gymnasium framework. The environment was constructed around the learned hydraulic model introduced in Section~\ref{sec:model_learn}.

Our goal is to develop a policy $\pi$ that adjusts the PI gains of the baseline feedback linearization (FL) controller~\eqref{eq:q24} to improve tracking performance. This approach follows the trend of recent works applying RL to enhance simple PID controllers for position control~\cite{yao2023data,tang2024reinforcement}. The policy outputs gain corrections as:

\begin{equation}
\pi(o_t) = [\Delta K_p, \Delta K_i],
\label{eq:policy_action}
\end{equation}

\noindent resulting in the modified FL control law:

\begin{equation}
u_{nom} = \frac{\dot{f_{r}} + (K_{p} + \Delta K_p)e + (K_{i} + \Delta K_i)\int e \, dt}{g} - \frac{h}{g}.
\label{eq:fl_rl}
\end{equation}

The observation vector provided to the agent extends the state $x_t$ with additional information relevant to tracking:

\begin{equation}\label{eq:obs}
o_t = [x_t, f_{r,t}, \dot{f}_{r,t}, e_t, \pi_{t-1}],
\end{equation}

\noindent where $f_{r,t}$ and $\dot{f}_{r,t}$ are the reference force and its derivative, $e_t = f_{r,t} - f_{l,t}$ is the tracking error, and $\pi_{t-1}$ is the previous policy action.

The reward function penalizes both squared error and error rate, strongly discouraging oscillatory or chattering behavior:

\begin{align}\label{eq:reward}
r(o_t,\pi_t) = -Q_1 e_t^2 - Q_2 \dot e_t^2.
\end{align}

\subsection{Agent Architecture and Training Setup}

The neural network architecture used for both actor and critic networks is intentionally compact to reduce computational overhead in real-time experiments, in contrast to larger networks typically employed in similar applications~\cite{sambhus2023real}. The main training hyperparameters and FL controller parameters are summarized in Table~\ref{tb:training_agent}.

\begin{table}[b]
\centering
\caption{SAC training and FL controller parameters.}
\label{tb:training_agent}
\small
\resizebox{0.8\linewidth}{!}{%
\begin{tabular}{ll|ll}
\toprule
\textbf{Parameter} & \textbf{Value} & \textbf{Parameter} & \textbf{Value} \\
\midrule
\rowcolor{lightgray}
Episodes                & $160$      & Steps per episode   & $1800$ \\
Actor layers            & $2$        & Actor layer size    & $32$  \\
\rowcolor{lightgray}
Actor activation        & ReLU       & Critic layers       & $2$   \\
Critic layer size       & $32$       & Critic activation   & ReLU  \\
\rowcolor{lightgray}
Actor LR                & $0.001$    & Critic LR           & $0.001$ \\
Initial $\mathcal{H}$   & $0.005$    & $\mathcal{H}$ LR    & $0.001$ \\
\rowcolor{lightgray}
$\gamma$                & $0.99$     & $\tau$              & $0.005$ \\
$Q_1$                   & $100$      & $Q_2$               & $4000$ \\
\rowcolor{lightgray}
$\Delta K_p^{max}$      & $75$       & $\Delta K_p^{min}$  & $0$ \\
$\Delta K_i^{max}$      & $10$       & $\Delta K_i^{min}$  & $0$ \\
\rowcolor{lightgray}
$K_p$                   & $15$       & $K_i$               & $5$ \\
\bottomrule
\end{tabular}}
\end{table}

\subsection{Real-World Traning Implementation}

Real-hardware training was carried out on the test bench hardware platform. A dSpace MicroLabBox handled low-level valve actuation and sensor data acquisition, while a personal computer equipped with an AMD Ryzen 5600G CPU and 16~GB RAM executed the SAC algorithm in Python. The PC received state measurements from the MicroLabBox, performed policy inference and training, and transmitted updated gain corrections back to the controller. Communication between the two systems was implemented via a serial protocol. Also, to speed up the training of the agent we began by loading an agent trained exclusively in simulation using our trained hydraulic model using the same parameters from Table \ref{tb:training_agent}.

To meet real-time constraints, the FL control loop was executed at 1000~Hz, while policy updates in~\eqref{eq:fl_rl} were applied at 100~Hz due to the additional computation required for inference and neural network learning. Training was carried out in randomized sinusoidal force trajectories, with amplitudes between 200–300~N and frequencies between 0.5–2~Hz. The inference of neural networks that were necessary for the execution of the training, such as the hydraulic model described in Section \ref{sec:model_learn} and the contraction metric, which will be described in \ref{sec:ccm}, was implemented by embedding the weights of the networks in the dSpace MicroLabBox industrial controller. The QP contraction filter described in Section \ref{sec:ccm} was implemented along with the controller described in \eqref{eq:fl_rl} with a value of $\lambda = 0.1$ to enforce stability during training and reduce vibration in the system. 

\section{Experiments}
\label{sec:results}

To validate the performance of our trained policy, we will evaluate the perforce of our agent with regard to the force tracking capacities analyzing the root mean squared error (RMSE) of the error in force tracking. We will compare our agent trained in the real-life system (Controller I) with another agent trained only in simulations using our trained hydraulic model (Controller II) described in Section \ref{sec:model_learn} using the same hyperparameters shown in \autoref{tb:training_agent} and an FL controller with fixed gains (Controller III) as described in \eqref{eq:q24} with values of $K_p = 90$ and $K_i =15$, which are the max values that the trained agent can deploy in the PI gains. The performance of the controllers was analyzed with reference signals of sin waves with 300N of amplitude with frequencies ranging from 0.5 Hz to 2 Hz. Table \ref{tb:rmse} shows the results.

\begin{table}[H]
\centering
\caption{Force tracking RMSE (N)}
\label{tb:rmse}
\begin{tabular}{|c|llll|}
\hline
\multicolumn{1}{|l|}{}          & \multicolumn{4}   {c|}{Frequency (Hz)}
\\ \hline
\multicolumn{1}{|l|}{} & \multicolumn{1}{c|}{0.5}  & \multicolumn{1}{c|}{1}  & \multicolumn{1}{c|}{1.5} & \multicolumn{1}{c|}{2}     \\ \hline
Controller I                              & \multicolumn{1}{l|}{5.5133} & \multicolumn{1}{l|}{5.9537} & \multicolumn{1}{l|}{8.7131} & 
\multicolumn{1}{l|}{11.3477} \\ \hline
Controller II                              & \multicolumn{1}{l|}{5.3214} & \multicolumn{1}{l|}{6.7550} & \multicolumn{1}{l|}{10.1796} & 
\multicolumn{1}{l|}{12.9672} \\ \hline
Controller III                              & \multicolumn{1}{l|}{3.2690} & \multicolumn{1}{l|}{6.3061} & \multicolumn{1}{l|}{9.7324} & 
\multicolumn{1}{l|}{12.9744} \\ \hline

\end{tabular}
\end{table}

It can be observed in Table \ref{tb:rmse} that the controller that used the agent trained in real time outperformed the controller that used an agent trained exclusively in simulation for every reference signal, with the exception of the 0.5 Hz signal, in which both controllers had very similar performance. These results demonstrate the advantages of applying the training of RL strategies in real-world systems, seeing that even high-precision models have inaccuracies that result in performance loss during real-world deployment. It can also be observed that Controller I outperformed the fixed-gain implementation (Controller III) for every reference signal with the exception of the 0.5 Hz one, demonstrating the potential of our implementation compared to more traditional control strategies.

\begin{figure}[htb]
  \centering
  \includegraphics[width=1\linewidth]{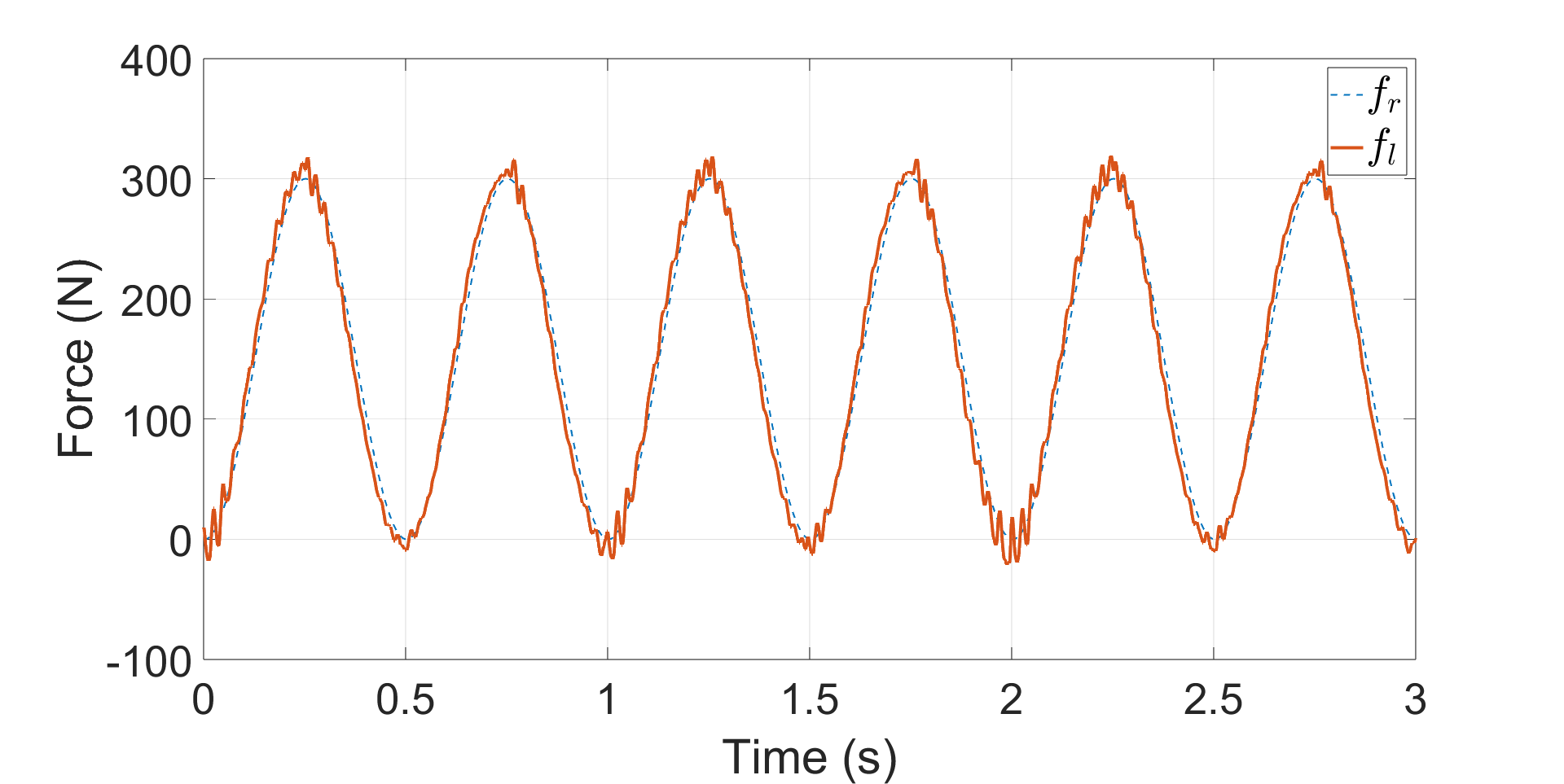}
  \caption{Force tracking for the 2hz experiments using Controller I} 
  \label{track_2hz}
\end{figure}

\begin{figure}[htb]
  \centering
  \includegraphics[width=1\linewidth]{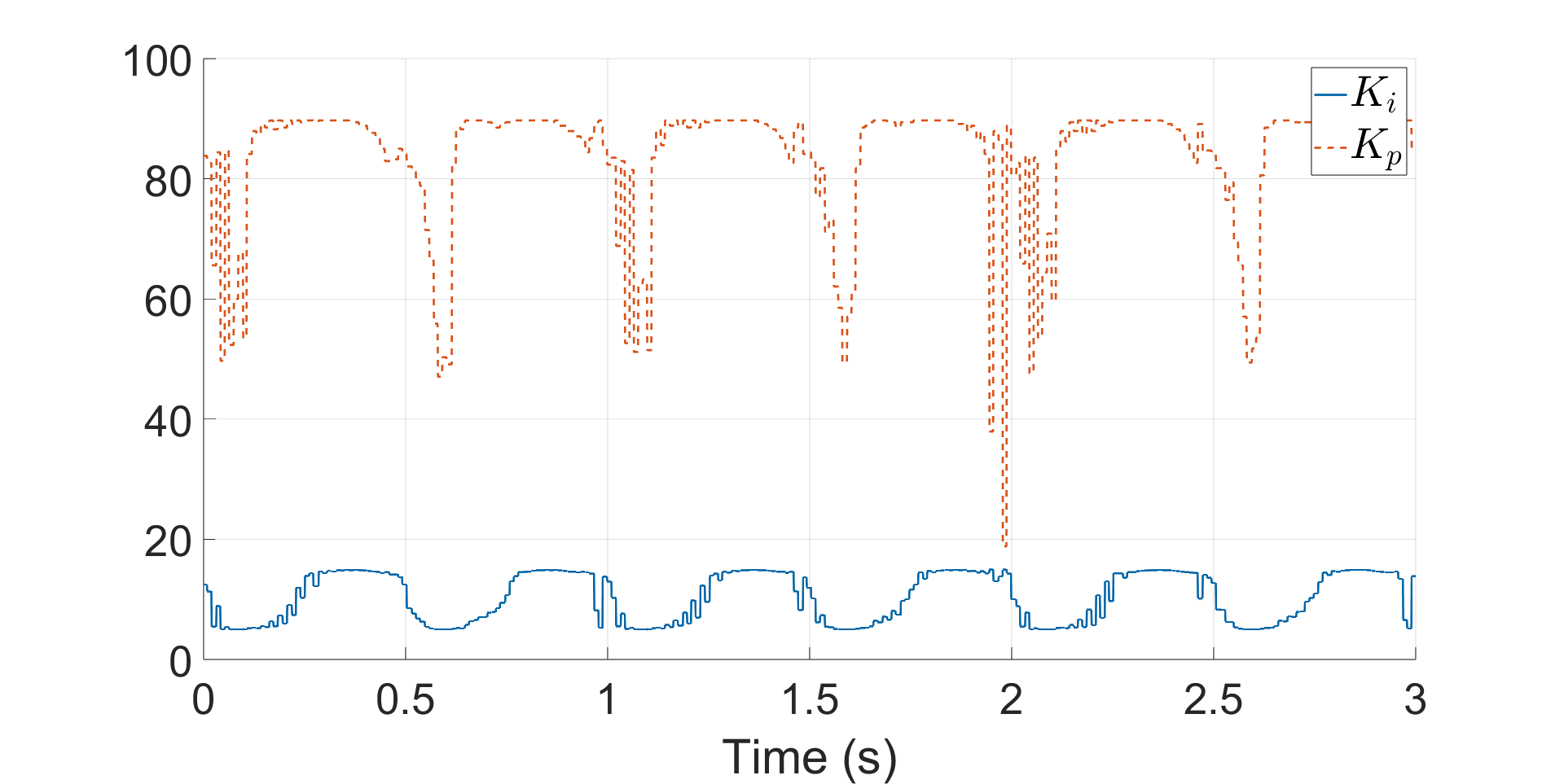}
  \caption{$K_p$ an $K_i$ variation using Controller I for the 2hz experiment} 
  \label{kp_kl_2hz}
\end{figure}

The performance of the real-life trained agent (Controller I) can be observed in Figure \ref{track_2hz} and Figure \ref{kp_kl_2hz} demonstrates how the agent varied the PI gains of \eqref{eq:fl_rl} in real time. The variation of the gains in real time was relatively smooth, ensuring that excessive vibrations didn't occur during controller deployment.

\section{Conclusion and future developments}
\label{sec:conc}
This paper introduced a contraction-metric-based framework for safe reinforcement learning in hydraulic force control. We first demonstrated how a supervised neural network model of the actuator dynamics can enable accurate simulation and policy pretraining. We then proposed a novel quadratic programming contraction filter, supported by a learned contraction metric, to enforce approximate contraction conditions in real time during hardware policy training. The resulting framework allowed us to safely train a Soft Actor-Critic policy directly on the hydraulic test bench, achieving superior tracking performance compared to both simulation-only training and fixed-gain feedback linearization.

The experiments confirm three main contributions: (i) the first demonstration of real-time RL training with a contraction-metric based safety filter on a hydraulic system; (ii) the application of contraction metrics to force control in hydraulics; and (iii) a practical QP-based contraction filter that enforces stability certificates online without compromising control performance. 

 While our approach significantly improves safety and performance, challenges remain. In particular, when controller gains become extreme, the filter shows sensitivity to measurement noise, partly due to limited noise exposure during metric training. Future directions will include extending the framework to multi-degree-of-freedom hydraulic systems and legged robots.


\bibliographystyle{IEEEtran}
\bibliography{references}

\end{document}